\documentclass[fleqn,10pt]{wlscirep}
\usepackage[utf8]{inputenc}
\usepackage[T1]{fontenc}
\usepackage{caption}
\usepackage{subcaption}
\usepackage{amssymb}
\usepackage{soul}
\usepackage[sorting=none,style=ieee]{biblatex}
\newcommand{\red}[1]{#1}
\newcommand{\delete}[1]{}
\newcommand{\mathred}[1]{#1}

\title{Unseeded One-Third Harmonic Generation in Optical Fibers}

\author[1,*]{Wen Qi Zhang}
\author[1]{Zane Peterkovic}
\author[1,2]{Stephen C. Warren-Smith}
\author[1]{Shahraam Afshar V.}
\affil[1]{Laser Physics and Photonic Devices Laboratories, UniSA STEM, University of South Australia, Mawson Lakes, South Australia 5095, Australia}
\affil[2]{Future Industries Institute, University of South Australia, Mawson Lakes, South Australia 5095, Australia}

\affil[*]{wenqi.zhang@unisa.edu.au}

\begin{abstract}
We propose a new concept to generate efficient one-third harmonic light from an unseeded third harmonic process in optical fibers. Our concept is based on the dynamic constant (Hamiltonian) of the nonlinear third harmonic generation in optical fibers and includes a periodic array of nonlinear fibers and phase compensation elements. We test our concept with a simulation of the nonlinear interaction between the fundamental and third harmonic modes of a realistic optical fiber, demonstrating high-efficiency one-third harmonic generation. Our work opens a new approach to achieving the so far elusive one-third harmonic generation in optical fibers.     
\end{abstract}
\begin{document}

\flushbottom
\maketitle

\thispagestyle{empty}

\section*{Introduction}

Third and one-third harmonic generation, THG and OTHG respectively, are third-order nonlinear processes in which three photons at a fundamental frequency convert to a photon at the third harmonic frequency and vice versa. The OTHG process, also known as third-order parametric down-conversion, has attracted interest due to the quantum properties of triple photon generation, e.g. entanglement \cite{gonzalez_continuous-variable_2018}. In pursuit of demonstrating OTHG, many theoretical and experimental approaches have been proposed and attempted. However, no successful demonstration of OTHG in waveguides has been reported. On the other hand, direct THG has been successfully demonstrated in bulk media and optical fibers with very low efficiency.

Despite the vast theoretical and experimental investigations into OTHG \cite{grubsky_glass_2005,grubsky_phase-matched_2007,huang_efficient_2013,evans_multimode_2015,moebius_efficient_2016,cavanna_hybrid_2016,warren-smith_direct_2016,jiang_fundamental-mode_2017,jiang_optimized_2018,huang_photonplasmon_2018, hammer_dispersion_2018,jiang_enhanced_2019,li_simultaneous_2020,cavanna_progress_2020,cheng_supercontinuum-induced_2020}, this process has not yet obtained practical use due to its low efficiency. Producing THG/OTHG in bulk materials is very inefficient due to a phase mismatch between the fundamental and third harmonic waves and a small interaction length. In optical waveguides, it is possible to achieve phase matching between the fundamental and higher-order spatial modes of the two different frequency waves for subwavelength waveguide geometries~\cite{grubsky_glass_2005}. Some recent reports have focused on the use of microstructured optical fibers pumped with a pulsed laser source \cite{warren-smith_third_2016}\cite{cheng_supercontinuum-induced_2020}\cite{grubsky_phase-matched_2007}. These have succeeded in producing THG with efficiencies on the order of $<0.01$\%, which is insufficient for achieving observable OTHG. As such, schemes exploiting the nature of certain key aspects, especially that of the phase-match condition, have been presented; these include schemes for tuning the phase-matching condition by controlling the gas pressure in hollow-core fibers \cite{nold_pressure-controlled_2010,cavanna_progress_2020} or surrounding submicron fiber tapers \cite{hammer_dispersion_2018}, tuning the phase matching through the application of coatings~\cite{warren-smith_nanofilm-induced_2017}, and counter-propagating beams for quasi-phase matching \cite{jiang_fundamental-mode_2017}.
However, no experimental demonstration of highly efficient THG (>1$\%$) or OTHG has been reported. The primary reason for this is that previous reports on an optimized geometry for THG/OTHG focused only on one parameter, namely phase matching between the fundamental fiber (waveguide) mode at one-third harmonic and higher-order modes at the third harmonic wavelengths\cite{grubsky_glass_2005,grubsky_phase-matched_2007,huang_efficient_2013,evans_multimode_2015,moebius_efficient_2016,cavanna_hybrid_2016,jiang_optimized_2018,huang_photonplasmon_2018, hammer_dispersion_2018,jiang_enhanced_2019}. It is apparent from the equations that govern the dynamics of THG and OTHG in optical waveguides (Eq. \ref{eq:nse}) that the THG process can grow from an unseeded process, while the inverse process cannot.

\delete{The theoretical studies on the THG and OTHG in optical fibers, usually indicate lengths of a few meters to achieve measurable efficiencies. In reality, the phase-matching condition cannot hold for such a long propagation length due to (i) the dynamic nonlinear phase, (ii) the nonuniformity of the waveguide geometry (the phase-matching condition is very sensitive to geometric/diameter fluctuations in subwavelength waveguides) and (iii) the propagation loss.} 

In our previous work, we have shown that THG and OTHG are multiparameter processes, which have stable and unstable solutions and high efficiency that can only be obtained in a stimulated process (seeded process) in the unstable regime \cite{afshar_v_efficient_2013}. 
In that regime, an interplay between the initial phase, the seed power, the phase mismatch, and different nonlinear coefficients determines the efficiency for THG and OTHG. Two dimensionless parameters, $(a, b)$ quantify the efficiency of THG or OTHG in optical fibers/waveguides, where $a\approx b\approx 0$ can lead to high efficiency THG or OTHG. While parameter $b=-(3\gamma_1+\gamma_3-3\gamma_{13}-\gamma_{31})/2\gamma_c$ is totally determined by the fiber/waveguide geometry, through the different effective nonlinear coefficients $\gamma_1,\gamma_3,\gamma_{13},\gamma_c$ (associated with third and one-third harmonics, and cross terms), parameter $a=(\gamma_3-3\gamma_{13})/\gamma_c+\delta\beta/P_0\gamma_c$, depends on the total input power $P_0$, as well as the fiber/waveguide structure, through the effective nonlinear coefficients $\gamma_1,\gamma_3,\gamma_{13},\gamma_c$ and the phase mismatch term $\delta\beta$ \cite{afshar_v_efficient_2013}.
In fibers (waveguides), it is usually difficult to simultaneously satisfy both the phase matching condition and the $a\approx b\approx 0$ condition for high efficiency. Note that the low efficiency in previous experimental reports supports our hypothesis. \delete{In addition, the fact that the phase-matching condition can not be held for the whole propagation length due to the change of nonlinear phase and the nonuniformity of the fiber geometry, also adds to the low efficiency of THG/OTHG processes.} \red{Our results in \cite{afshar_v_efficient_2013} indicates that high-efficiency OTHG is possible in the seeded regime when the pump and the seed lasers at fundamental and third harmonic frequencies are phase-locked. This, however, cannot be readily achieved among two different laser sources, especially considering the requirement for synchronisation in time, in the case of pulsed lasers, and launching the pump and seed into specific modes.}


In this paper, we propose an experimentally viable scheme based on the theory work in \cite{afshar_v_efficient_2013} to achieve high-efficiency OTHG through a \red{cascaded approach} initiated from an unseeded process. \red{Contrary to the seeded process, the unseeded one requires no initial phase-locking condition, time synchronisation, and launching into a specific higher-order spatial mode. Hence, our proposed approach relaxes critical practical conditions necessary for high-efficiency OTHG.} Considering a Hamiltonian of the THG and OTHG processes, we propose a periodic array of fibers and phase compensator pairs to achieve high-efficiency THG and OTHG. 
The phase compensator element after every fiber allows the required linear phase shift between the interacting waves (fundamental and third harmonic) in order to achieve an efficiency increase after every pair. Our idea is similar to the idea of periodically poled lithium niobate (PPLN) waveguides for high-efficiency second harmonic generation \cite{myers_periodically_1997}. In PPLN a periodic array of pairs of nonlinear crystals with opposite crystal orientations is constructed to maintain the phase matching (quasi-phase matching) condition throughout the waveguide. The concept of our idea is based on our theoretical work in \cite{afshar_v_efficient_2013}, which was limited to THG/OTHG processes in ideal lossless, dispersionless fibers without any Raman effect. We test our idea for realistic fibers by developing a  numerical simulation of THG and OTHG processes in an optical fiber including losses (insertion losses), dispersions, and Raman effects. The fiber parameters are extracted based on the scanning electron microscope (SEM) image of a realistic microstructured optical fiber that was used in an experiment demonstrating THG \cite{warren-smith_exposed_2014}. We show that THG/OTHG efficiency higher than $30\%$ is possible in this fiber. 


\section{Concept}
\label{concept}In this section, we develop the main theory behind our PFPC scheme. 

\begin{figure}[h]
    \centering
    \includegraphics[height=6cm]{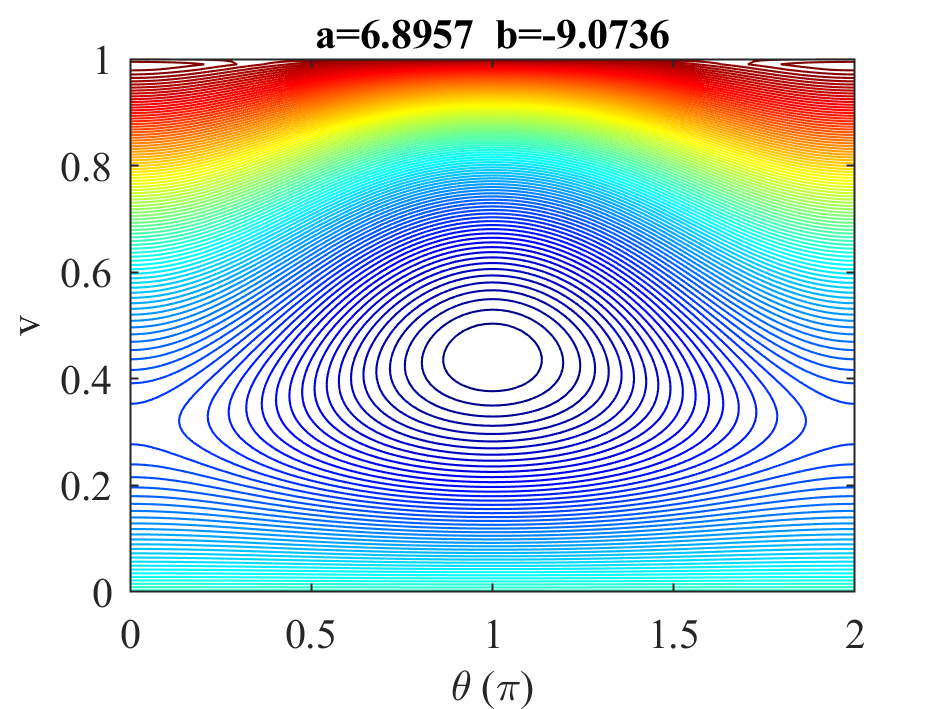}
    \includegraphics[height=6cm]{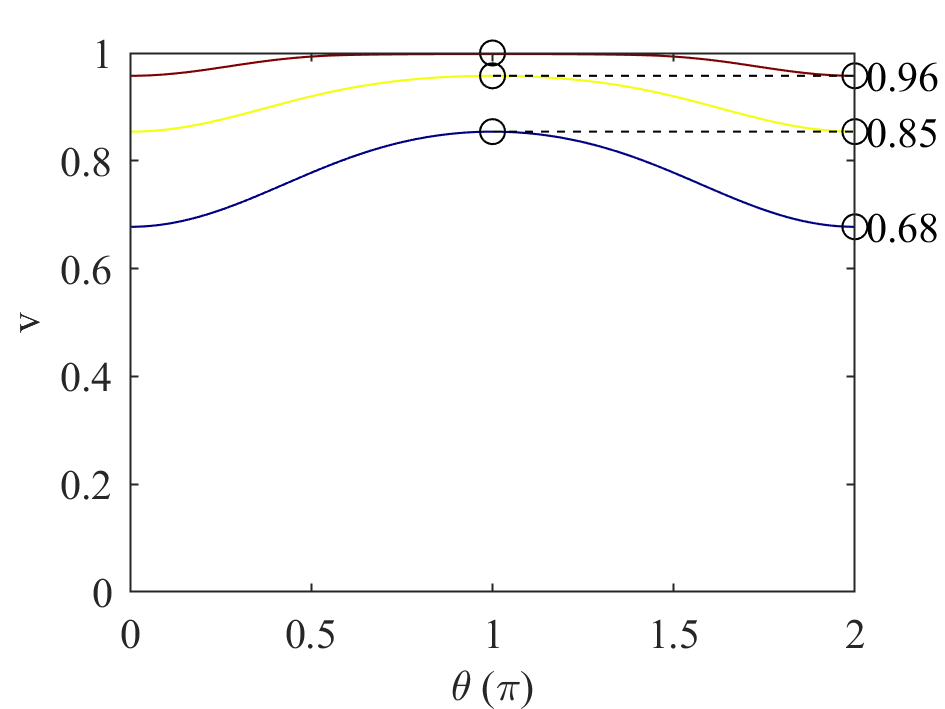}
    \caption{Left: constant Hamiltonian contour map as a function of $v$ and $\theta$ for a fiber with $(a,b)=(6.8957,-9.0736)$. For initial values of $(v_0,\theta_0)$, $v$ and $\theta$ follow one of the contours as laser light propagates through the fiber. Right: our phase compensation concept. Starting with an initial shallow contour (brown curve) and applying $\pi$ phase shift (dotted lines) allows accessing deeper contours with higher conversion efficiencies $\Delta v$ (yellow and blue curves)  }
    \label{fig:H}
\end{figure}

THG and OTHG in optical waveguides can be described by the following coupled equations: 
\begin{eqnarray}
\frac{\partial}{\partial z}A_1\mathred{+}\alpha_1 A_1\mathred{-i}\sum_{n=1}^{\infty} \frac{\beta_1^{(n)}}{n!}(\frac{\mathred{i}\partial}{\partial t})^nA_1&=&i\left(\gamma_1\left|A_1\right|^2+i\gamma_{13}\left|A_3\right|^2\right)A_1+i\gamma_c A_1^{\ast 2}A_3 e^{i\delta\beta z},\nonumber\\
\frac{\partial}{\partial z}A_3\mathred{+}\alpha_3 A_3\mathred{-i}\sum_{n=1}^{\infty} \frac{\beta_3^{(n)}}{n!}(\frac{\mathred{i}\partial}{\partial t})^nA_3&=&i\left(\gamma_3\left|A_3\right|^2+i\gamma_{31}\left|A_1\right|^2\right)A_3+i\gamma_c A_1^3 e^{-i\delta\beta z},\label{eq:Fnse}
\end{eqnarray}
where $A_1$, $A_3$ are the slowly varying amplitudes of the fundamental and third harmonic frequencies for different spatial modes, respectively, $\delta\beta=\beta_3(3\omega)-3\beta_1(\omega)$ is the detuning of phase match, $\gamma_1$, $\gamma_2$, $\gamma_{13}$,$\gamma_{31}$ and $\gamma_c$ are nonlinear coefficients \red{associated with self-phase modulation ($\gamma_1$, $\gamma_2$), cross-phase modulation ($\gamma_{13}$, $\gamma_{31}$), and four-wave mixing ($\gamma_c$)} and can be calculated based on the modal overlaps between the pump and third harmonic waves (see Appendix \ref{nonlinear_coefficients}), $\alpha_{1,3}$ are fiber losses associated with fundamental and third harmonic frequencies, respectively, and $\beta_{1,3}^{(n)}$  are different dispersion orders for the fundamental and third harmonic fields, respectively. Equations~\ref{eq:Fnse} can be simplified to
\begin{eqnarray}
\frac{\partial}{\partial z}A_1&=&i\left(\gamma_1\left|A_1\right|^2+i\gamma_{13}\left|A_3\right|^2\right)A_1+i\gamma_c A_1^{\ast 2}A_3 e^{i\delta\beta z}\nonumber\\
\frac{\partial}{\partial z}A_3&=&i\left(\gamma_3\left|A_3\right|^2+i\gamma_{31}\left|A_1\right|^2\right)A_3+i\gamma_c A_1^3 e^{-i\delta\beta z},\label{eq:nse}
\end{eqnarray}
in the steady state limit \red{(ignoring dispersion)}, and for lossless fibers. These equations can be rewritten in terms of power ratio $v$ and phase difference $\theta$:
\begin{eqnarray}
\frac{\partial}{\partial\tau}v&=&-2v\sqrt{v(1-v)}\sin\theta\nonumber\\
\frac{\partial}{\partial\tau}\theta&=&a+2bv+(4v-3)\sqrt{\frac{v}{1-v}}\cos\theta\label{eq:vth}
\end{eqnarray}
where $v=\mathred{P_1}/P_0$, $\theta=\phi_3-3\phi_1+\delta\beta z$, $\phi_1$ and $\phi_3$ are the phase of $A_1$ and $A_3$ respectively, $\tau=\gamma_c P_0 z$, $\mathred{P_1=|A_1|^2}$, $\mathred{P_3=|A_3|^2}$,  $P_0=\mathred{P_1+P_3}$, $a=(\gamma_3-3\gamma_{13})/\gamma_c+\delta\beta/P_0/\gamma_c$ and $b=-(3\gamma_1+\gamma_3-3\gamma_{13}-\gamma_{31})/2/\gamma_c$. The maximum conversion efficiency of THG or OTHG can be fully parameterized by $a$ and $b$: $a$ and $b$ having opposite signs and relatively small values (preferably smaller than 10) can lead to high efficiency \cite{afshar_v_efficient_2013}. Equations \ref{eq:vth} can be obtained from a Hamiltonian, $H(\nu,\theta)$,  
\begin{equation}
 H(v,\theta)=-av^2+2v\sqrt{v(1-v)}\cos{\theta},
 \label{eq:H}
\end{equation}
which is the constant of the THG-OTHG process, $\partial H/\partial \tau=0$. 
This means that for an initial condition of $(v_0,\theta_0)$, the evolution of THG/OTHG process,
$v(\tau)$, and $\theta(\tau)$, inside a waveguide follows one of the contours $H(v,\theta)$, 
which is constant for all values $v$ and $\theta$. 
An example of the contour map of different constant Hamiltonian is shown in Fig. \ref{fig:H} for $(a,b)=(6.8957,-9.0736)$.  

\begin{figure}[h]
    \centering
    \includegraphics[width=10cm]{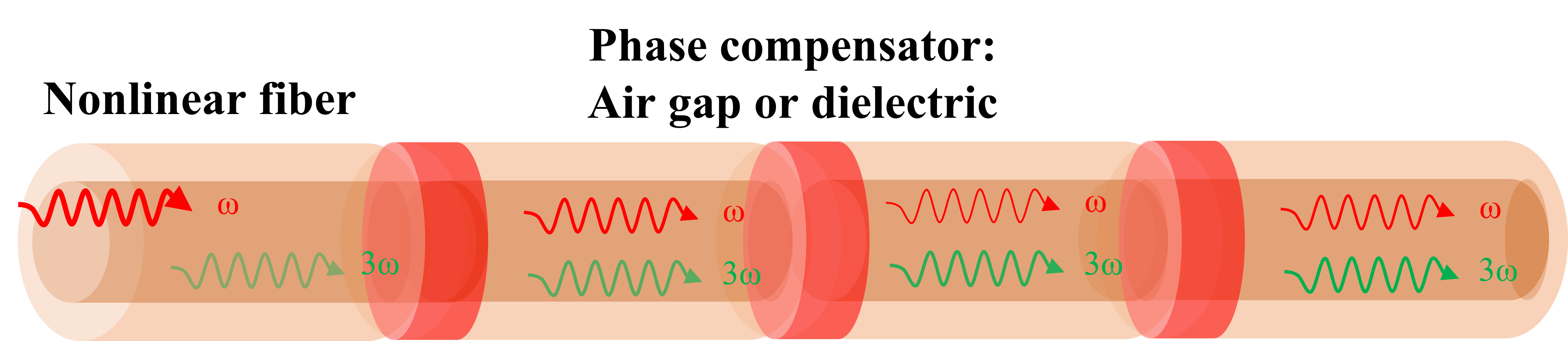}
    \caption{Schematic of the main concept: a periodic array of nonlinear optical fiber segments separated by phase compensator regions (\delete{air gaps or } e.g., dielectric coatings) which allow the required phase shift compensation between the third and one-third harmonic waves before entering the next fiber segment. The process is unseeded initiating from a pump wave at the fundamental wavelength of $\lambda=1552.8$nm.}   
    \label{fig:expsetup}
\end{figure}

The contour map of constant H, Fig. \ref{fig:H}, is the main concept in developing our new phase compensation idea. It is evident by the contour that processes with $v_0\approx 0$ ($P_3\approx P_0$) stay close to $v\approx 0$ contours regardless of their $\theta_0$ values (i.e., no unseeded OTHG generation), however, processes with $v_0\approx 1$ ($P_1\approx P_0$) follow a contour that enables small reduction in $v$, i.e. unseeded THG generation. 
This is consistent with Eqs. \ref{eq:nse} that indicates the possibility of generation of $A_3$ from an initial $A_1\neq 0, A_3=0$ (unseeded THG) but not the reverse process: generation of $A_1$ from an initial $A_3\neq 0, A_1=0$ (unseeded OTHG). As a result, OTHG is only possible through a seeding process, requiring an initial fixed phase difference, $\theta_0$, between the pump, $A_3$, and the seed, $A_1$ waves. 
This requirement is hard to achieve and maintain pulse to pulse, considering the input pump and seed lasers need to be configured at different spatial modes (usually pump at higher order and seed at the fundamental spatial modes). Hence, no OTHG has been observed experimentally yet. 

Our proposed scheme is as follows: we start with an unseeded THG ($v_0\approx0$ shallow contour in Fig. \ref{fig:H} right), which has low conversion efficiency but generates a phase-locked pump and seed laser at the right spatial modes, and apply a series of phase compensations on the pump and seed to access deeper contours with higher conversion efficiencies. This concept is shown schematically in Fig. \ref{fig:H} right, in which applying $\pi$ phase shift \red{in $\theta=\phi_3-3\phi_1$} enables accessing the third contour (blue curve), which has the conversion efficiency of $\Delta \nu \approx 32\%$. The phase compensation can be achieved by having bulk dielectric coatings between segments of optical fibers \delete{ or simply air gaps}, as shown in Fig. \ref{fig:expsetup}. The idea of phase compensation here is similar to the quasi-phase-matching $\chi^{(2)}$ nonlinear applications such as second harmonic generation (SHG), where multiple SHG processes are stacked up by periodically adjusting the phase between pump and signal~\cite{myers_periodically_1997}.

\section{Numerical model}
While the theoretical model in the previous section explains the concepts behind our phase compensation scheme, it is far from a real experimental situation. For instance, equations \ref{eq:nse} are developed for CW laser beams with no dispersion, no loss, and no walk-off between the third and one-third harmonic generations. In addition, it has been assumed that multiple phase compensations between different fiber segments are lossless. In this section, we develop a numerical model to explore the above concept for realistic experimental parameters. The model includes the THG-OTHG generation in exposed core fibers, Sec. \ref{THG-OTHG}, and field propagation within phase compensation regions, Sec \ref{Phase}.

\subsection{THG-OTHG in exposed core fibers}
\label{THG-OTHG}
We consider an exposed core silica microstructured optical fiber used previously to demonstrate direct THG with a femtosecond laser pump \cite{warren-smith_third_2016}. \red{This exposed-core fiber was fabricated in-house at the University of Adelaide from pure silica glass (F300, Heraeus, $n = 1.4446$ at $\lambda = 1500 nm$ and $n = 1.4623$ at $\lambda = 500 nm$) as previously described in \cite{kostecki_predicting_2014}. It has an outer diameter of 145 $\mu m$ and an effective core diameter of 1.85 $\mu m$ (Fig.~\ref{fig:SEM}, Left). The calculated nonlinear coefficients (based on equations in Appendix \ref{nonlinear_coefficients}) of this fiber are: $\gamma_1=0.0619$ W$^{-1}$m$^{-1}$, $\gamma_3=0.1651$ W$^{-1}$m$^{-1}$, $\gamma_{13}=0.0442$ W$^{-1}$m$^{-1}$, $\gamma_{31}=0.1325$ W$^{-1}$m$^{-1}$ and $\gamma_c=0.0047$ W$^{-1}$m$^{-1}$.} In that study, a maximum THG output of 5.3 $\mu$W with a 49.3 mW pump power was achieved indicating a conversion efficiency of approximately 0.01\%. We have taken the scanning electron microscope image of the exposed-core fiber (shown in Fig.~\ref{fig:SEM}, Left) and used it together with the Sellmeier equation of the silica glass~\cite{malitson_interspecimen_1965} to model the fiber eigenmodes at the pump and TH wavelengths using COMSOL Multiphysics. \red{Adaptive meshes were used in COMSOL to ensure the effective refractive indices of modes were converged to $10^{-6}$.}
\begin{figure}[h]
    \centering
    \includegraphics[height=6cm]{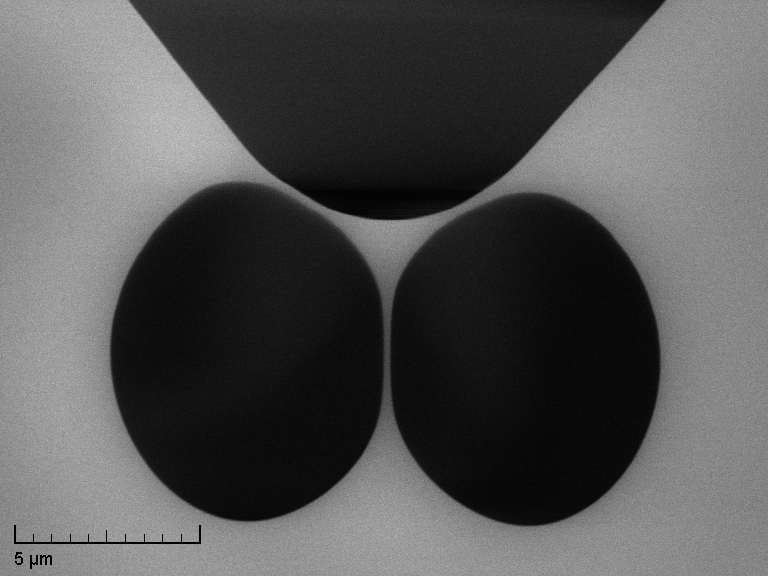}
    \includegraphics[trim=0.4cm 0.25cm 0.6cm 0.35cm, clip=true,height=6cm]{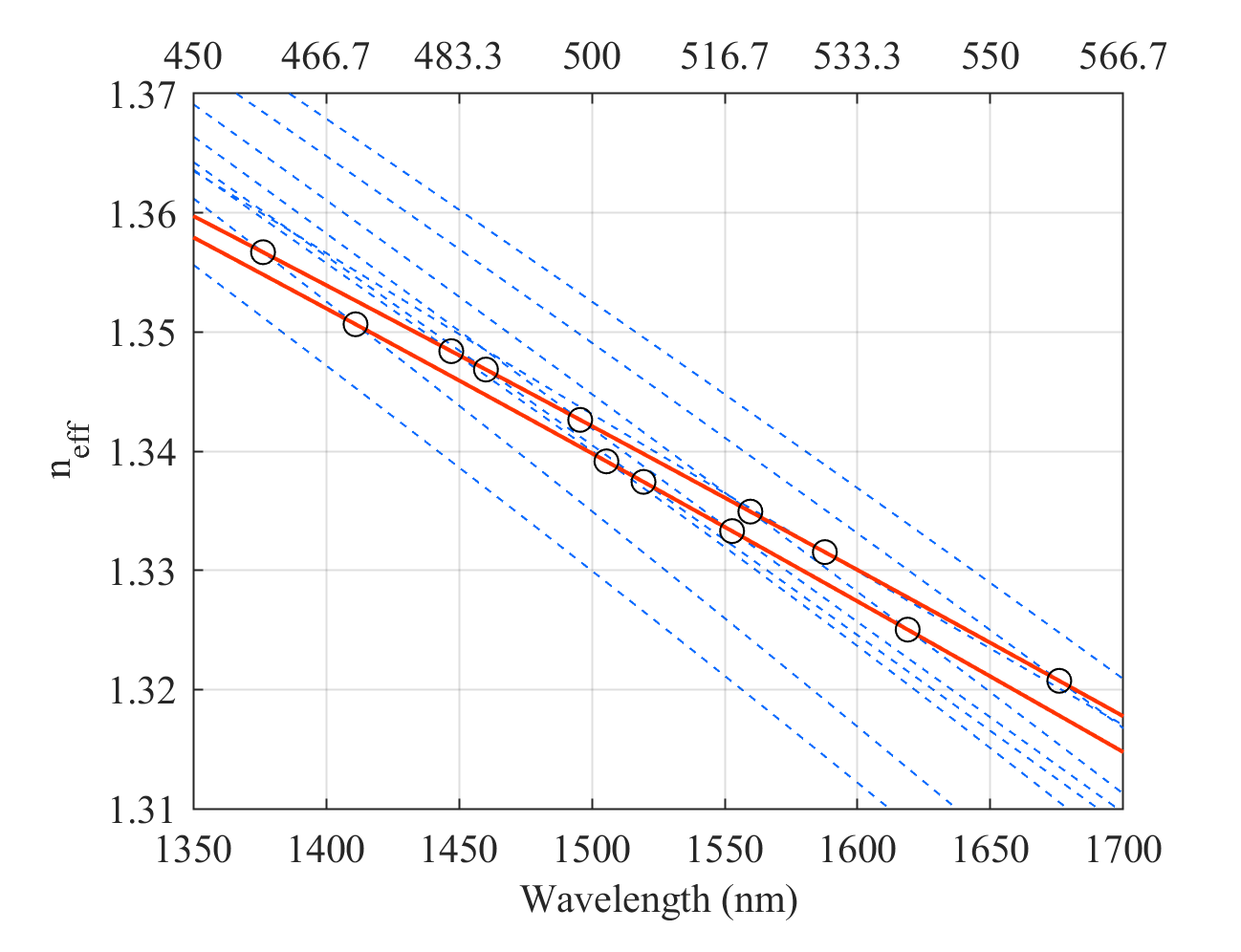}
    \caption{ Left, scanning electron microscope image of the cross-section of the exposed core fiber used in our previous THG experiment \cite{warren-smith_third_2016}, and used as the basis of the numerical simulations in this work. Right, the effective refractive indices of the fundamental modes (two polarisations) in the range of 1350-1700 nm (solid red lines), and higher order modes in the range of 450-566.7 nm (dashed blue lines). The circles show the phase-matched points.}
    \label{fig:SEM}
\end{figure}
The effective refractive indices ($n_\text{eff}$) of the pump modes (solid curves) and THG modes (dotted curves) are shown in Fig.~\ref{fig:SEM}, Right. \delete{THG} \red{Zero-detuning phase match points ($\delta\beta =0$)} can be found at various wavelengths across different modes (depicted as circles in the plot). It is worth noting that there is an uncertainty in determining the exact position of the edges of the fiber geometry from the SEM image, which could lead to discrepancies between the simulation and experiment, especially affecting the phase-matched modes.

\begin{figure}[h]
    \centering
    \includegraphics[width=12cm]{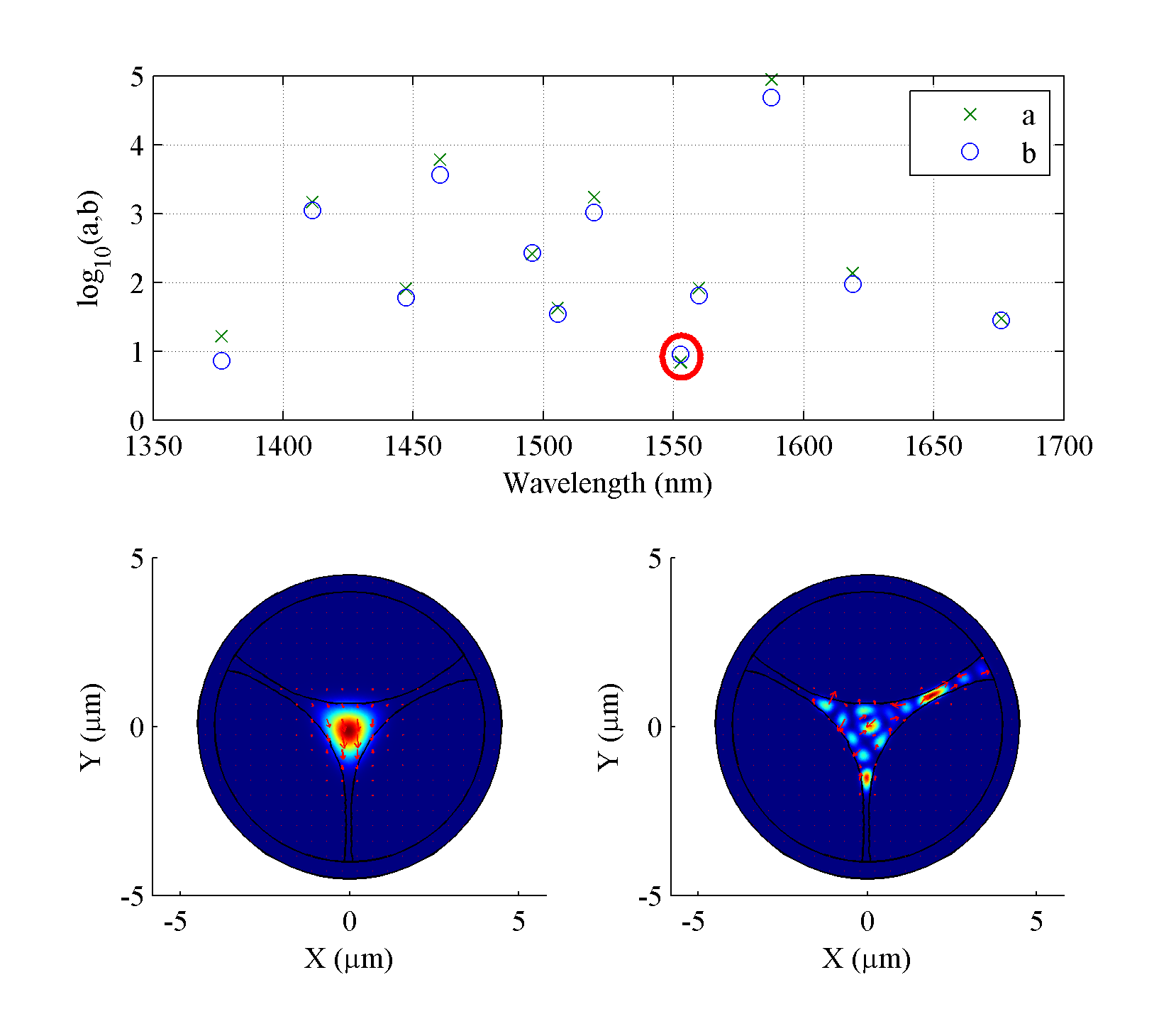}
    \caption{Top: $a$ and $b$ values \red{(log$_{10}$ values}) for the phase-matched modes identified in Fig. \ref{fig:SEM}. The red circle corresponds to  $a=6.8957$ and $b=-9.0736$, for which the fundamental and TH higher spatial modes (power flow distribution) are shown in the bottom left and bottom right, respectively. All the simulations have been carried out based on these two modes.}
    \label{fig:ab}
\end{figure}

COMSOL was used to calculate the nonlinear coefficients and the $a$ and $b$ parameters for all the phase-matched points in Fig.~\ref{fig:SEM} with a total peak power $P_0=1000$ W; these have been plotted in Fig.~\ref{fig:ab} (top). The smallest $a$ and $b$ are found around 1552.8 nm (circled in red) for a pair of modes with $a=6.8957$ and $b=-9.0736$. This pair of modes and their associated $a$ and $b$ parameters are used throughout the simulation and in plotting the Hamiltonian in Fig. \ref{fig:H}. The power distributions of the pair of modes at the pump (bottom, left) and TH (bottom, right) wavelengths are also plotted in Fig.~\ref{fig:ab}.

In Fig. \ref{fig:H}, starting from an initial value of  $v_0\approx 1$, the Hamiltonian trajectory can reach a minimum $v$ of 0.94, which indicates a theoretical maximum THG conversion efficiency of $\Delta \nu \approx 6\%$. This value assumes the pump is a single-frequency (CW) laser with 1000 Watts of peak power. As the linewidth of the pump laser increases, the efficiency reduces. To achieve high peak power, pulsed laser pumps are usually required, for which the efficiency will be lower than 6\%. Especially for femtosecond pumps, where significant pulse broadening will take place and other nonlinear processes such as self-phase-modulation (SPM), cross-phase-modulation (XPM), and simulated Raman scattering (SRS) will also have significant impacts on the THG efficiency. Furthermore, for femtosecond and picosecond pulsed laser pumps, the walk-off between the pump and THG pulses due to the difference in group velocities limits the maximum interaction length, which further limits the THG efficiency. To achieve the best THG performance we focus on nanosecond pulsed lasers. 
We consider an input hyperbolic secant pulse for the fundamental wave 
$$
q(t)=A\cdot\text{sech}(t),
$$
where $A$ is the amplitude of the pulse and hence at any time instance, $P_1(t)=P_0(t)=|q_0(t)|^2$. The initial $v_0$ and $\phi_0$ of different time instances across the pulse are the same at the input to the fiber, however they, as well as $P_0$, change during propagation. Therefore, the phase compensation optimization idea, discussed in the Sec. \ref{concept}, only works for a single point in time, i.e. the center of the pulse at $t=0$, and is used as a guide to estimate the required phase compensation. Furthermore, the difference in the group velocities between the pump and THG wavelengths leads to a walk-off in time. Although for nanosecond pulses, the walk-off over 10s of centimeters of fiber is not as significant as picosecond and femtosecond pulses, it limits the maximum propagation length over which the CW model applies.

To account for all parameters affecting THG, we perform a numerical simulation based on Eqs. \ref{eq:Fnse} for the spatial modes that are shown in Fig. \ref{fig:ab}. These equations account for all dispersion terms, losses, SPM, and XPM for the two spatial modes at the fundamental and third harmonic frequencies as they propagate along the fiber. The numerical calculation of Eqs.~\ref{eq:Fnse} is based on a split-step Fourier method (SSFM) where the nonlinear step is solved using a 4th-order Runge-Kutta method~\cite{agrawal_nonlinear_2001, hult_fourth-order_2007}. Since with SSFM, the dispersion terms of Eqs.~\ref{eq:Fnse} are solved in the frequency domain, we included the full dispersion profile of the modes from COMSOL instead of using Taylor expansion. \red{The step size of the pulse propagation simulation was set to 10$\mu m$ with a temporal window of 200 $ns$ to ensure the stability and convergence of the results.} \red{We have considered a loss of 0.1 dB/m for 1552.8 nm and 0.3 dB/m for 517.6 nm, which is typical for these fiber structures. Although, we have noticed that fiber loss doesn't impact the overall results significantly, see Fig. \ref{fig:NSpulse90T2} (Middle).} The nonlinear coefficients of the modes used in the numerical simulation are also obtained from COMSOL \red{as explained in the Appendix \ref{nonlinear_coefficients}}.

\subsection{Phase compensation}
\label{Phase}
In this section, we consider \delete{the free-space (air gap) or }bulk (dielectric coating) propagation of fiber modes outside the fiber and their coupling into the next fiber segment, consistent with our proposal for phase compensation as schematically shown in Fig. \ref{fig:expsetup}.
To model the propagation of different modes outside the fiber, we \red{consider the general solution to the paraxial wave equation and} apply the following equation to the electric and magnetic fields of the fundamental and TH waves:
$$
\bold{e}(x,y,z+d)=F^{-1}[e^{-i(\frac{2\pi^2}{2kn}(\nu^2+\mu^2)d)}F[\bold{e}(x,y,z)]],
$$
where $\bold{e}(x,y,z)$ is the electric field distribution of the mode, $\mathred{d}$ \red{is the length of the phase compensation region (dielectric coating) between two fibers}, $F$, $F^{-1}$ are the Fourier and inverse-Fourier transforms, respectively, $n$ is the refractive index of the phase compensation medium, and $\nu,\mu$, are the spatial frequency variables \cite{okamoto_fundamentals_2006}. Depending on the length of the phase compensation region, its refractive index, wavelengths of the fundamental and TH waves, and their spatial mode profiles, a phase shift is induced between the fundamental and TH waves in propagating through the phase compensation region.  After propagating through the phase compensation region \delete{air gap}, fundamental and TH signals are coupled into the next fiber segment. 

The coupling coefficient between the fundamental and TH spatial modes and the modes of the next fiber segment can be \red{approximately calculated by ignoring the Fresnel reflection and using the overlap integrals}: 

$$
a_i = \frac{1}{2}\int_{\infty} \bold{e}_i^{inc} \times \bold{h}_i^*\cdot\hat{z} ~dA,
$$

in which $a_i$ is the complex amplitude ($|a_i|^2$, therefore, is the coupling efficiency), $\bold{e}_i^{inc}$ is the incident electric field vector, and $\bold{h}_i$ is the magnetic field of the same mode as $\bold{e}_i^{inc}$ in the next fiber. Both $\bold{e}_i^{inc}$ and $\bold{h}_i$ are normalized to the mode power. \red{To compensate for Fresnel reflection,  we have multiplied the coupling coefficient for each frequency by the Fresnel transmission coefficient} $t=\displaystyle \frac{4 n_1 n_2}{(n_1+n_2)^2}$, where $n_1$ is the refractive index of the dielectric coating and $n_2$ is the effective refractive index of the mode. The complex amplitude $a_i$ carries the phases acquired by the fundamental and TH waves in propagating through the phase compensation region.

\section{Results}
\label{Results and discussion}
\begin{figure}[h!]
    \centering
    \includegraphics[width=10cm]{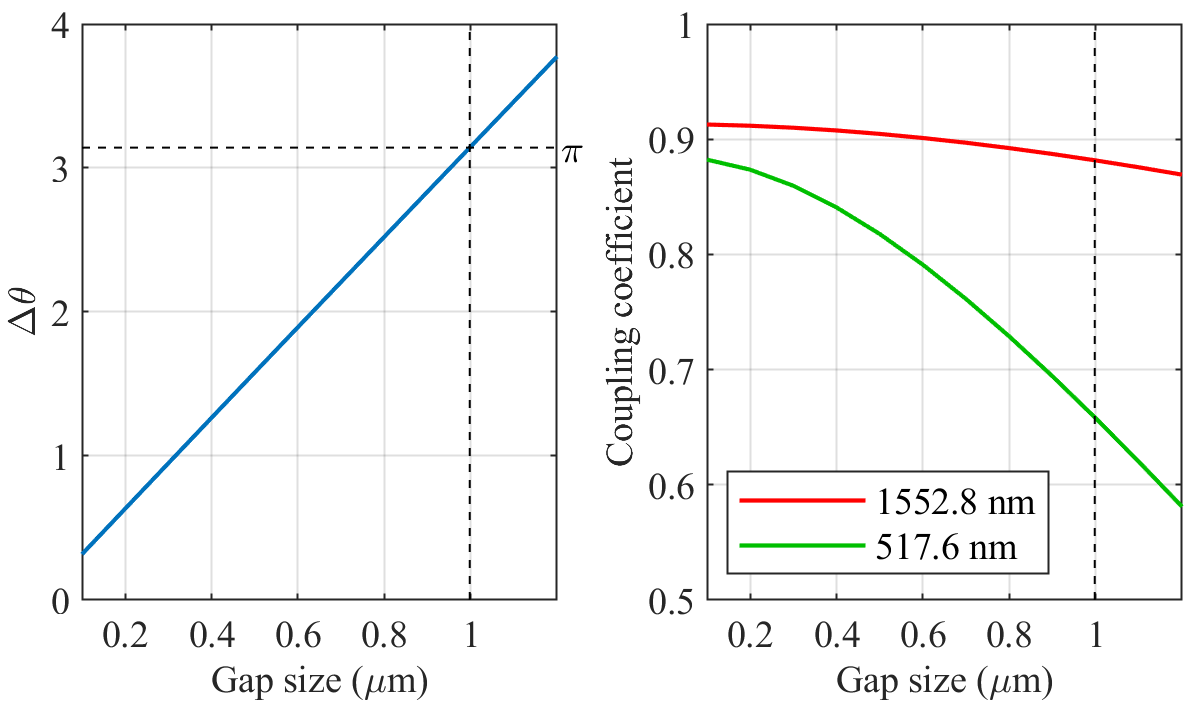}
    \caption{Left: phase shift in $\theta=\phi_3-3\phi_1$ as a function of ZnSe gap length. Right: the magnitude of the coupling coefficients of the fundamental and TH modes, red and green respectively, to the similar modes in the next fiber segment as a function of \delete{air gap} \red{dielectric coating gap size}.}
    \label{fig:CouplingEff}
\end{figure}

We consider an unseeded THG process (see Fig. \ref{fig:expsetup}) for which the input pump is a 7 ns pulse at the wavelength of 1552.8 nm with a peak power of 1 kW. The input is coupled to the fundamental mode of the fiber and the TH is generated in a higher order mode, whose profiles are shown in Fig. \ref{fig:ab} (bottom, left, and right, respectively).  We also consider \delete{air gaps} \red{Zinc Selenide (ZnSe), as the dielectric coating} between fiber segments for phase compensation regions (see Fig. \ref{fig:expsetup}) and show the phase shift \delete{between the fundamental and TH modes} \red{in $\theta$, i.e., $\Delta\theta$} as a function of gap length in Fig. \ref{fig:CouplingEff} (left).
The coupling coefficients for the fundamental spatial mode and TH higher order spatial mode as a function of \delete{air gap} \red{dielectric coating thickness} are also shown in Fig. \ref{fig:CouplingEff} (right). The TH signal shows lower coupling efficiency than the fundamental signal due to increased diffraction of higher-order spatial modes. \red{Zinc Selenide (ZnSe) has a high dispersion between 517.6 nm to 1552.8 nm spectra range to allow phase compensation with a short propagation length ($\sim1\mu m$).}

The discrepancy in coupling efficiencies of the TH and the fundamental leads to a different input value of $v$ for a fiber segment compared to the $v$ of the previous fiber segment. Overall, each \delete{air gap} \red{dielectric coating gap} produces a shift in the input $\theta$ and $v$ of each fiber segment compared to those values at the end of the previous fiber segment. Based on the theoretical solutions of Eqs. \ref{eq:vth} and \ref{eq:H}, Fig \ref{fig:H} (right) suggests $\pi$ phase shift \delete{between the fundamental and third harmonic} \red{in $\theta$} to access contours of constant Hamiltonian with higher efficiencies. Hence, in our numerical simulation (based on solutions of Eqs. \ref{eq:Fnse}), we adopt a strategy to have a $\pi$ phase shift \red{in $\theta$} for every \delete{air gap} \red{dielectric coating gap} in our simulation (by keeping the length of the dielectric coating gap constant), and optimize the length of the subsequent fiber for the highest THG or OTHG efficiency. This strategy has been shown to lead to the highest THG efficiency.   

\begin{figure}[h!]
    \centering\includegraphics[width=14cm]{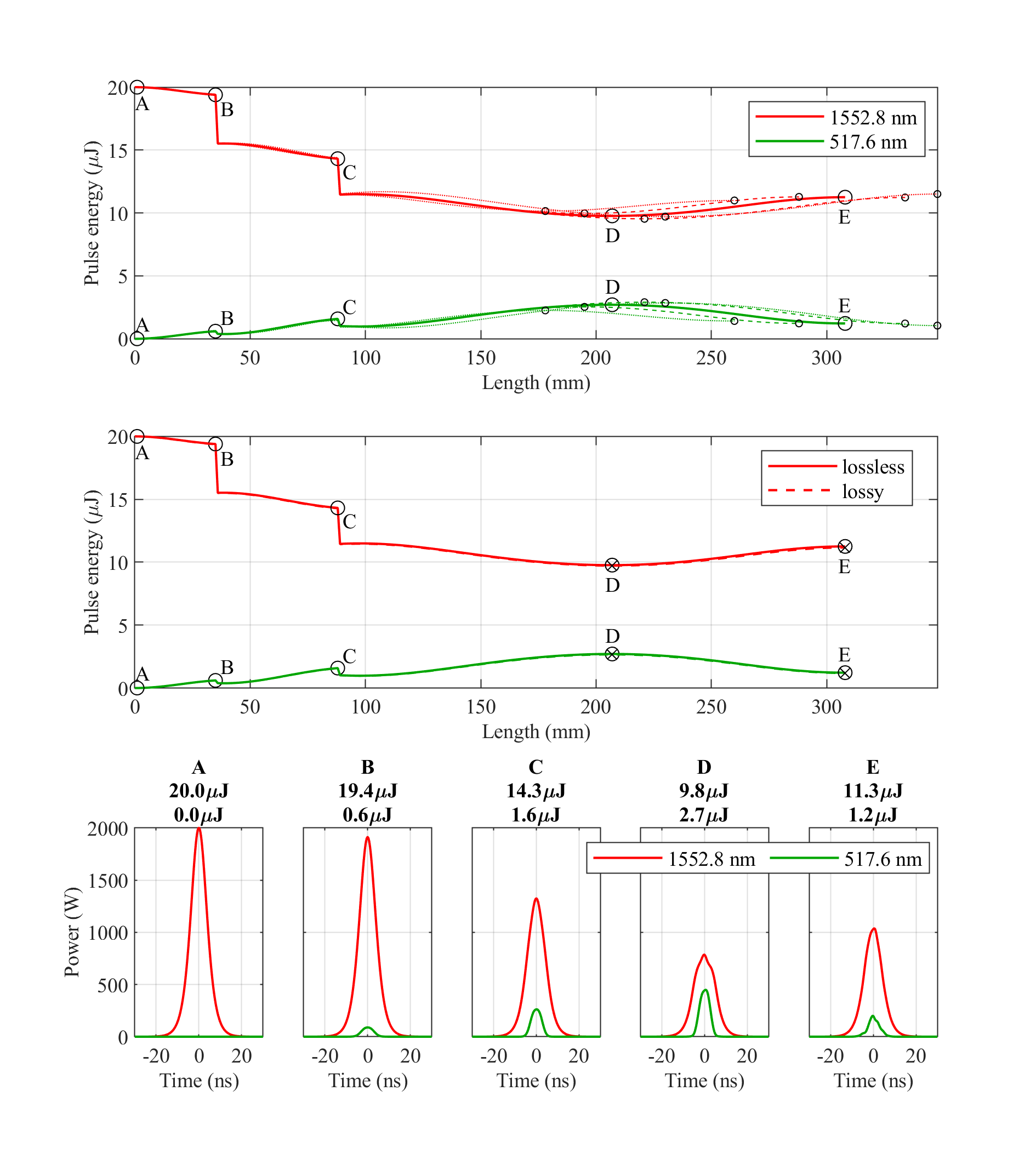}
    \caption{(Top) The fundamental (red) and TH pulse (green) energies as a function of length along the periodic array of fiber-dielectric coating gap geometry (three fiber segments and two dielectric coating gaps (ZnSe)). Sharp energy drops represent the loss in the dielectric coating gaps. \red{The dashed and dotted curves correspond to $\pm5$\% variations in $a$ and $b$, respectively.} \red{(Middle) Comparison of models with (solid curves) and without (dashed curves) propagation loss, assuming a 0.1 dB/m loss for 1552.8 nm and a 0.3 dB/m loss for 517.6 nm.} (Bottom) The temporal profiles of the fundamental (red) and TH (green) waves at the beginning (A) of the fiber-dielectric coating array, and the end of the first (B), second (C), and third segments (E). (D) is the maximum THG point in the third segment. OTHG in the last segment yields a \red{54.9}\% conversion of the TH pulse energy back into the pump pulse and a \red{13.2}\% energy increase of the pump pulse compared to point (D).}
    \label{fig:NSpulse90T2}
\end{figure}

Using the above strategy for choosing the \delete{air gap} \red{dielectric coating gap} lengths, we have considered an array configuration that consists of \delete{six} \red{three} fiber segments and \delete{five air gaps} \red{two dielectric coating gaps}. The length of each segment has been chosen to maximize THG in that segment. 
Figure \ref{fig:NSpulse90T2}, top and bottom show the pulse energy and the pulse temporal profile, respectively, of the initial pump source (point A) and output of the TH and fundamental signals at the end of the \delete{second} \red{first} (point B), \delete{fourth} \red{second} (point C), and \delete{sixth} \red{third} segments (point E). 
As a general observation, the generated THG pulses are narrower in time than the pump pulse. As the pulses propagate through the fiber segments, the pump pulse has deformed significantly due to depletion.

Comparing the pulse energies for the fundamental and TH at the end of each segment, Fig.~\ref{fig:NSpulse90T2} (Top) shows how these energies change progressively over fiber segments. The drop in pulses' energies at the end of each segment is due to the losses in \delete{air gaps} \red{dielectric coating gaps}. 
In the \delete{sixth} \red{third} segment, the maximum TH energy is achieved around 225 mm, (point D) See Fig. ~\ref{fig:NSpulse90T2} (Top). This is the maximum achievable TH signal energy for the chosen fiber configuration in this work and we have noticed small changes in the TH pulse energy beyond the \delete{sixth} \red{third} segment hence we have stopped adding more \delete{air gaps} \red{dielectric coating gaps} and fiber segments. This is due to the total loss of power in the TH and fundamental signals and the temporal walk-off between them, which reduces their nonlinear interaction. Considering the initial pump and the maximum TH pulse energies of 20 $\mu$J and 2.7 $\mu$J, respectively, we have achieved around 13.5$\%$ conversion efficiency.

In the \delete{sixth} \red{third} segment, we have considered elongation of the segment to allow the conversion of the TH signal to the fundamental frequency, i.e. an OTHG process. This can be seen in Fig.~\ref{fig:NSpulse90T2} (top, point D to E), in which the fundamental pulse energy increases by the end of the \delete{sixth} \red{third} segment around 350 mm (point E). The temporal profiles of both TH and fundamental signals at the end of the \delete{sixth} \red{third} segment are also shown in Fig.~\ref{fig:NSpulse90T2} (bottom). Considering pulse energies in the \delete{sixth} \red{third} segment, we find the conversion efficiency from the TH to the fundamental signal, i.e., OTHG process, to be around \red{$54.9\%$}. \red{For comparison, the theoretical maximum conversion efficiency for $a=6.8957$ and $b=-9.0736$ in a single segment is $55.0\%$.} It is also apparent that \red{$13.5\%$} of the final fundamental signal energy is obtained through the TH $\rightarrow$ OTH process compared to the rest of the fundamental signal energy which is the residue of the input pump. This is fundamentally important since the \red{13.5$\%$} energy in the fundamental signal is due to the fission of a high energy photon at $\lambda$=517.6 nm into three photons at $\lambda$=1552.8 nm. The three generated photons are theorized to be entangled \cite{gonzalez_continuous-variable_2018} and hence have different quantum properties compared to the other photons in the fundamental signal. As such, one would expect to be able to demonstrate the quantum properties of the generated triple entangled photons. Alternatively, our results indicate that a pump pulse of \red{20} $\mu$J energy can generate \red{1.49} $\mu$J of triple entangled photons, which is equivalent to \red{$7.4\%$} conversion efficiency. We expect such efficiency can be improved by optimizing the fiber design and increasing the input pump power.

\red{To evaluate the influence of structural variation and laser power fluctuation on conversion efficiency, we have added $\pm$5\% variation to $a$ and $b$. The parameters $a$ and $b$ are independently tuned through $\delta\beta$ and $\gamma$s. According to the definition of $a$, changing $\delta\beta$ has the same effect as changing $P_0$. We have re-run the simulation for the new $a$ and $b$ values and kept the lengths of the first and second segments the same but varied the length of the third segment to search for the maximum and minimum power in $P_1$ and $P_3$. The results for variations of a and b have been plotted as dashed and dotted curves in Fig.\ref{fig:NSpulse90T2}(top), respectively. The corresponding OTHG conversion efficiencies in the last segment are 51.8\% ($1.05a$), 58.2\% ($0.95a$), 63.2\% ($1.05b$) and 37.3\% ($0.95b$). The percentages of energy in the pump pulse due to OTHG are 11.6\% ($1.05a$), 15.1\% ($0.95a$), 15.6\% ($1.05b$) and 7.7\% ($0.95b$).}

\section{Conclusion}

We have proposed an approach to achieve one-third harmonic generation (third-order parametric down conversion) in optical fibers through an unseeded third harmonic generation. Our approach is based on a periodic array of nonlinear fibers and phase compensators. Fibers enable the nonlinear interaction between the fundamental and third harmonic waves while the phase compensators allow certain phase shifts between the two waves. In this way, the array can be engineered to achieve a high-efficiency conversion of energy from one wave to the other wave. We have tested our approach by developing a realistic simulation model of the array and shown that for parameters of real fibers (loss, nonlinearity, dispersion), we can achieve \delete{32$\%$} \red{54.9$\%$} conversion efficiency in THG$\rightarrow$OTHG. We believe this provides a practical approach to achieving OTHG in optical fibers, something that has yet to be demonstrated. Considering the quantum properties of OTHG in optical fibers, i.e. the entanglement of the generated triple photons, our approach would provide a practical way to achieve a fiber-based quantum entanglement source for quantum applications.

\section*{Acknowledgements}
This research was supported fully by the Australian Government through the Australian Research Council (DP190102896). Stephen C. Warren-Smith is supported by an Australian Research Council Future Fellowship (FT200100154).



\printbibliography

\section{Appendix}
\label{nonlinear_coefficients}

Effective nonlinear coefficients associated with the fundamental and third harmonic frequencies in a fiber (waveguide) can be derived from Maxwell's equations considering a third-order nonlinear polarization $P_{NL}^{(3)}$. In a multimode fiber, the effective nonlinear coefficients of different spatial modes associated with the fundamental and third harmonic frequencies can be complex \cite{hartley_cross_2015}, however, if we restrict the interaction to two modes: fundamental spatial mode at the fundamental frequency and a higher order mode at the third harmonic frequency then the nonlinear coefficients can be given as:  
\begin{align}\label{gammavariables}
    \begin{split}
    \gamma_1 &= \frac{2\pi\epsilon_0}{3\mu_0\lambda_1}\int n_1^2n_{(2)}[2|\bold{e_1}|^4 + |\bold{e_1}^2|^2]dA  \\
    \gamma_3 &= \frac{2\pi\epsilon_0}{3\mu_0\lambda_3}\int n_3^2n_{(2)}[2|\bold{e_3}|^4 + |\bold{e_3}^2|^2]dA \\
    \gamma_{13} &= \frac{4\pi\epsilon_0}{3\mu_0\lambda_1}\int n_1n_3n_{(2)}[|\bold{e_1}|^2 \cdot |\bold{e_3}|^2 + |\bold{e_1}\cdot \bold{e_3}|^2+|\bold{e_1}\cdot \bold{e_3}^{\ast}|^2]dA  \\
    \gamma_{31} &= \frac{4\pi\epsilon_0}{3\mu_0\lambda_3}\int n_1n_3n_{(2)}[|\bold{e_1}|^2 \cdot |\bold{e_3}|^2 + |\bold{e_1}\cdot \bold{e_3}|^2+|\bold{e_1}\cdot \bold{e_3}^{\ast}|^2]dA \\
    \gamma_{c} &= \frac{2\pi\epsilon_0}{3\mu_0\lambda_1}\int n_1^{\frac{3}{2}}n_3^{\frac{1}{2}}n_{(2)}[(\bold{e_1}^{\ast}\cdot \bold{e_3})(\bold{e_1}^{\ast}\cdot \bold{e_1}^{\ast})]dA.
    \end{split}
\end{align}

Here, $\bold{e_{1,3}}$ are the modal electric field distribution for the fundamental (fundamental spatial mode) and third harmonic (higher order spatial mode) frequencies, respectively, $n_{1,3}$ are refractive indices of silica at the respective frequencies, $n_{(2)}$ is the nonlinear refractive index of silica, $n_{(2)}\approx2.2\times10^{-20}$ m$^2W^{-1}$ \cite{agrawal_nonlinear_2001}. Electric field distributions $\bold{e_{1,3}}$ are normalized to $\displaystyle \sqrt{N_{1,3}}$, where $\displaystyle N_{1,3}=\frac{1}{2}\int_{\infty} (\bold{e_{1,3}} \times \bold{h^*_{1,3}}) \cdot \hat{z}dA$, respectively.


\end{document}